\documentclass[12pt]{iopart}

\usepackage{graphicx}

\usepackage{amssymb}

\newcommand{\diag}{{\rm diag}}

\begin{document}

\title[Two-photon decays of vector mesons and dilepton decays  of scalar
mesons...]{Two-photon decays of vector mesons and
dilepton decays of scalar mesons in dense matter}

%\author{A E Radzhabov}\ead{aradzh@thsun1.jinr.ru}\,
%and \author{M.K.Volkov}\ead{volkov@thsun1.jinr.ru}
\author{A E Radzhabov,
M K Volkov and V L Yudichev}

\address{Bogoliubov Laboratory of Theoretical Physics, \\
Joint Institute for Nuclear Research, 141980 Dubna, Russia}
\eads{\mailto{aradzh@thsun1.jinr.ru},\mailto{volkov@thsun1.jinr.ru},
\mailto{yudichev@theor.jinr.ru}}

\begin{abstract}
Two-photon decays of vector mesons and dilepton decays of scalar
mesons which are forbidden in vacuum and
can occur in dense baryonic matter due to the explicit violation
of Lorentz symmetry are described within a quark model of the
Nambu--Jona-Lasinio type. The temperature and chemical potential
dependence of these processes is investigated. It is found that
their contribution to the  production of photons and leptons in
heavy-ion collisions is enhanced near the conditions corresponding
to the restoration of chiral symmetry.  Moreover, in the case of
the $a_0$ meson and especially the $\rho$-meson, a resonant
behaviour (an additional amplification) is observed due to the
degeneration of $\rho$ and $a_0$ masses when a hot hadron matter
is approaching  a chirally symmetric phase.
\end{abstract}

\submitto{\JPG}

%\noindent{\it Keywords\/}
%vector and scalar mesons, electromagnetic decays, dense baryonic matter

%\maketitle

\section{Introduction}
The medium-induced breaking of Lorentz symmetry for the ground
state of hadron matter, while the interaction  is
Lorentz-invariant, can open some processes that are forbidden in normal
conditions (in free space). The mixing of scalar and vector mesons
such as $\sigma$-$\omega$,  $\rho$-$a_0$ and $\phi$-$f_0(980)$ is
an example of this. The role of $\sigma$-$\gamma$  transition in
the enhancement of $e^+e^-$ production due to the (forbidden in
vacuum) decay $\sigma\to e^+e^-$ near the two-pion threshold was
discussed by Weldon in \cite{Weldon:1991ej}. (According to
vector-meson dominance (VMD), all electromagnetic interactions of mesons
should be mediated by vector mesons: a scalar transforms to a
vector which then transforms to a photon.) The effect of
$\sigma$-$\omega$ and $\rho$-$a_0$ mixing  on the pion-pion
 annihilation in dilepton production was investigated in
\cite{Teodorescu:2000mg}, where it was shown that additional peaks
appeared in the dilepton spectrum in a sufficiently dense baryonic
matter. On the other hand, the scalar-vector mixing can trigger
two-photon decays of vector mesons, where a vector meson first
transforms to a neutral scalar meson which then produces two
correlated photons. In the case of $\rho$-$a_0$ mixing, an
additional (resonant) amplification can occur because of the
degeneration of the $\rho$ and $a_0$ meson masses in particular
conditions in  dense matter \cite{Radzhabov:2005fh}.

The electromagnetic processes are considered as well-suitable
probes to study in-medium properties of mesons in heavy-ion
collisions. Much efforts have  been sofar directed toward the
understanding of meson properties in baryonic matter, because they
can strongly modify in hot and dense medium as a precursor
phenomenon of chiral symmetry restoration
\cite{Brown:1991kk,Brown:1995qt,Rapp:1999ej,Gazdzicki:2004}.
Insofar as correlated photon pairs are more difficult objects for
detection than charged leptons, the most of attention has been
paid to observation of dilepton spectra.
 Some measurements of dilepton spectra in heavy-ion collisions
have been carried out at Lawrence Berkeley National Laboratory
\cite{Porter:1998wy}, CERES \cite{Agakishiev:2005nm} at CERN, the
NA38 and NA50 Collaborations \cite{Abreu:1999av,Abreu:1999jr}.
Experiments are also planned by PHENIX at the Relativistic Heavy
Ion Collider at Brookhaven National Laboratory and by HADES at the
Gesellschaft f\"ur Schwerionenforschung (GSI)
\cite{Friese:1999qm}. The facility planned to be constructed at
GSI can produce more dense baryonic matter, which better suits for
the observation of the above mentioned processes as they are more
sensitive to density than to temperature. As to  low-energy
photon-photon correlations, some experimental and theoretical
opportunities of their observation at Nuclotron (Dubna) had been
envisaged recently during the round table discussion
"Searching for the mixed phase of strongly interacting matter at the JINR Nuclotron"
in June  2005 (see {\tt http://thsun1.jinr.ru/meetings/2005/roundtable/talks.html}).

The pion-pion  annihilation is usually considered as the main
mechanism of two-photon  and dilepton production (see e.~g.\
\cite{Anchishkin:2005yy}), while direct decays of mesons are of
low interest. Due to modification of scalar and vector mesons in
hot and dense medium, the spectrum of produced particles both in
the annihilation and in direct decays can change. Thus, possible
enhancement of photon pair production in the process
$\pi\pi\to\gamma\gamma$ due to modification of $\sigma$ was
investigated e.~g.\ in \cite{Volkov:1997dx,Volkov:2002bv}. A
peculiarity of the pion-pion annihilation via $\sigma$ is that its
enhancement is expected in a very narrow region in the phase
diagram. It is impossible  to maintain a system with fixed
temperature and chemical potential in heavy-ion experiments,
therefore the produced particles spectra should be averaged over
all phases through which the matter evolves during the collision.
Unfortunately, the background processes shadow the peak in the
two-photon enhancement because  the conditions relevant for the
enhancement exist for very short time. On the contrary, the
increase in two-photon or dilepton production from the processes
with $\sigma$-$\omega$, $\rho$-$a_0$ and $\phi$-$f_0$ mixing
survives in a more wide range of temperature and baryon density,
which improves the chances of their observation.

In our paper, we focus our attention on some electromagnetic and
strong decays of scalar and vector mesons that occur due to
$\sigma$-$\omega$, $\rho$-$a_0$ and $\phi$-$f_0$ mixing in dense
baryonic matter. In difference to \cite{Weldon:1991ej}, where only
pion-loop contributions were taken into account, we use  a
three-flavour version of the Nambu--Jona-Lasinio (NJL) model with
't~Hooft interaction in the mean-field approximation to estimate
the contribution of quark loops to the scalar-vector mixing,
because,  in the large-$N_c$ (number of colors) approximation,
quark loops give the leading contribution. The quark-loop
contribution to the dilepton production in pion-pion annihilation
has been already estimated in \cite{Blaschke:1997jj} for
deconfined quarks only. We calculate the partial widths
corresponding to direct decays of scalar and vector mesons in the
hadron phase but near the transition to the phase with restored
chiral symmetry.
 The "real-time"
formalism is implemented to introduce medium effects for the quark
propagator in quark-loop calculations
\cite{Dolan:1973qd,Ebert:1992ag}. The vector-meson dominance
is used for the description of dilepton decays of scalar mesons.

%
% Paper structure
%
The structure of our paper is as follows. In section~2, the model
Lagrangian introduced and parameters are given. Two-photon and
some strong decays of vector mesons $\rho$, $\omega$ as well as
two-photon decays of $\phi$ are described in section~3. In
section~4, dilepton decays of scalar mesons $\sigma$, $a_0$ and
$f_0(980)$ are calculated. In the last section, we draw our
conclusion and give discussion of the obtained results.

\section{Model and parameters}

For the description of meson properties, we use a $U(3)\times
U(3)$ NJL model with 't~Hooft interaction
\cite{Volkov:1986zb,Klimt:1989pm,Vogl:1991qt,Klevansky:1992qe,Volkov:1993jw,Ebert:1994mf}.
The Lagrangian of the model consists of two parts: a $U(3)\times
U(3)$  symmetric four-quark interaction and 't~Hooft determinant
\cite{'tHooft:1976up} with six-quark vertices
\begin{eqnarray}
\mathcal{L} &= {\bar q}(i{\hat \partial} - m^0)q
 + \frac{G_1}{2}\sum_{i=0}^8 \left[({\bar q}{\lambda}_i q)^2 +({\bar q}i{\gamma}_5{\lambda}_i q)^2\right]
 - \frac{G_2}{2}\sum_{i=0}^8 ({\bar q}\gamma_\mu{\lambda}_i q)^2 -\nonumber\\
&- K \left\{ {\det}[{\bar q}(1+\gamma_5)q]+{\det}[{\bar q}(1-\gamma_5)q] \right\}
\label{Ldet},
\end{eqnarray}
where $\lambda_i$ (i=1,...,8) are the Gell-Mann matrices and $\lambda^0 = {\sqrt{2\over
3}}${\bf 1}, with {\bf 1} being the unit matrix; $q=\{u,d,s\}$ and $\bar{q}$ are
quark and antiquark fields;  $m^0=\diag(m^0_{\rm u},m^0_{\rm d}\approx m^0_{\rm u},m^0_{\rm s})$
is the current quark mass matrix; $G_1$ and $G_2$ are the four-quark interaction
constants in the scalar-pseudoscalar and vector channels; $K$ is the six-quark interaction
constant.

Chiral symmetry is known to be spontaneously broken in vacuum,
which leads to the shift of the ground (vacuum) state to one with
a nontrivial chiral quark condensate. The latter entails the
formation of constituent masses of quarks related to the current
quark masses by gap equations. For three flavors with isotopic
symmetry, the gap equations are
\begin{eqnarray}
m_{\rm d}&=m_{\rm u}=m_{\rm u}^0 + 8 m_{\rm u} G_1
I_1^\Lambda(m_{\rm u})
	+32 m_{\rm u} m_{\rm s} K I_1^\Lambda(m_{\rm u}) I_1^\Lambda(m_{\rm s}),\label{gapu}\\
m_{\rm s}&=m_{\rm s}^0 + 8 m_{\rm s} G_1 I_1^\Lambda(m_{\rm s})+32
 K \left(m_{\rm u}I_1^\Lambda(m_{\rm u})\right)^2, \label{gaps}
\end{eqnarray}
where $m_{\rm u}$ and $m_{\rm s}$ are the constituent quark masses;
$I_1^\Lambda(m)$ is a quadratically
divergent integral that comes from the tadpole diagram and depends on the corresponding
constituent quark mass $m=\{m_{\rm u}, m_{\rm s}\}$ and
the  ultraviolet cut-off parameter $\Lambda$. In the three-dimensional
 regularization scheme it equals
\begin{equation}
\label{I1}
I_1^{\Lambda_3}(m)=\frac{N_c}{(2\pi)^2} \int\limits_0^{\Lambda_3}\frac{p^2}{E}\,d p.
\end{equation}
The subscript 3 at $\Lambda_3$ means that the cut-off is implemented in three-dimensional
momentum space.

The model parameters in vacuum are fixed in a way that allows
one to reproduce the masses of $\pi$, $K$ and $\rho$ mesons, the pion weak-decay constant
$f_\pi=92.4$~MeV, the strong decay width of the $\rho$-meson and
the mass difference of $\eta$ and $\eta'$ mesons (see \cite{PDG}).
We do not describe here the details of the parameterization scheme
(the reader will find it e.~g.\ in \cite{Volkov:1986zb}) but give the
following result
$m_{\rm u}=280$~MeV, $m_{\rm u}^0=2.1$~MeV,
$m_{\rm s}=416$~MeV, $m_{\rm s}^0=51$~MeV,
$G_1=3.2$~GeV$^{-2}$, $G_2=16$~GeV$^{-2}$,
$K = 4.6$~GeV$^{-5}$, $\Lambda_3 = 1.03$~GeV.

In hot and dense hadron gas, medium effects should be taken into account.
In the mean-field approximation, where only one-loop diagrams are calculated,
one can use the "real time" formalism \cite{Dolan:1973qd,Ebert:1992ag},
where the quark propagator is replaced by its extended form
\begin{eqnarray}
S(p,T,\mu)&=(\hat{p}+m)\left[ \frac{1}{p^2-m^2+i\epsilon}\right.\nonumber\\
& +2\pi i
\delta(p^2-m^2)(\theta(p^0)n(\mathbf{p};T,\mu)+\theta(-p^0)n(\mathbf{p};T,-\mu)) \Biggr],
\end{eqnarray}
with all medium effects being contained in the Fermi-Dirac
function for quarks
\begin{eqnarray}
n(\mathbf{p};T,\mu)=\left(1+\exp\frac{E-\mu}{T}\right)^{-1}.
\end{eqnarray}
Here, $E=\sqrt{\mathbf{p}^2+m^2}$, $p$ and $\mathbf{p}$ are four-
and three-momentum, respectively; $T$ is the temperature and $\mu$ is
the baryonic chemical potential. The hat over $\hat{p}$ stands for the
contraction with $\gamma$-matrices:
$\hat{p}=p_\alpha\gamma^\alpha$.

All divergent parts of the one-loop diagrams necessary for our
calculations can be represented
by two basic integrals with quadratic and logarithmic ultraviolet divergencies
\begin{eqnarray}\label{I1T}
I_1^{\Lambda_3}(m,T,\mu)&=& \frac{N_c}{(2\pi)^2}\int \limits_0^{\Lambda_3} dp
\frac{p^2}{E}\left(1-n(p;T,\mu)-n(p;T,-\mu)\right),\\
I_2^{\Lambda_3}(m,T,\mu)&= &\frac{N_c}{2(2\pi)^2}\int
\limits_0^{\Lambda_3} dp
\frac{p^2}{E^3}\left(1-n(p;T,\mu)-n(p;T,-\mu)\right).
\end{eqnarray}
In \eref{I1T}, the reader will recognize the thermalized form of
the integral $I_1^{\Lambda_3}(m)$ introduced in (\ref{I1}).

\begin{figure}[tbp]
\caption{The $u$- and  $s$-quark masses as functions of $\mu$ at
fixed temperature: $T=20$~MeV.}\label{Fig_qmasses}
\begin{center}
%{\includegraphics[scale=1]{mumsT=20.eps}}
\includegraphics[scale=1]{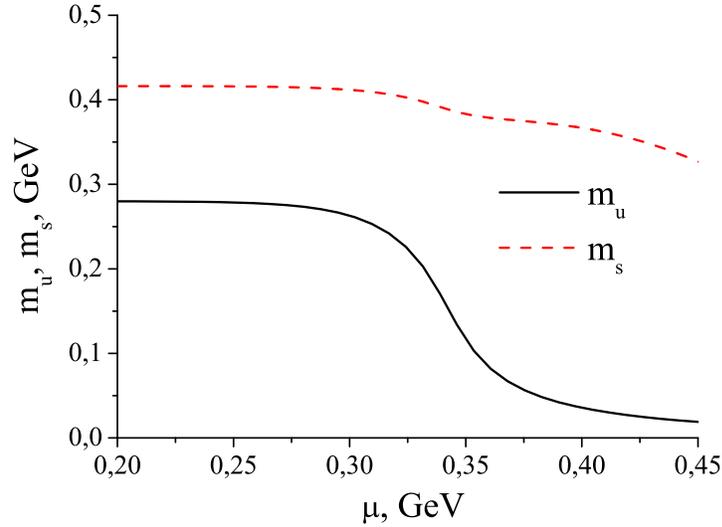}
\end{center}
\end{figure}

\begin{figure}[tbp]
\caption{The $u$- and  $s$-quark masses as functions of $\mu$ at
fixed temperature: $T=100$~MeV.}\label{Fig_qmasses_1}
\begin{center}
%{\includegraphics[scale=1]{mumsT=120.eps}}
{\includegraphics[scale=1]{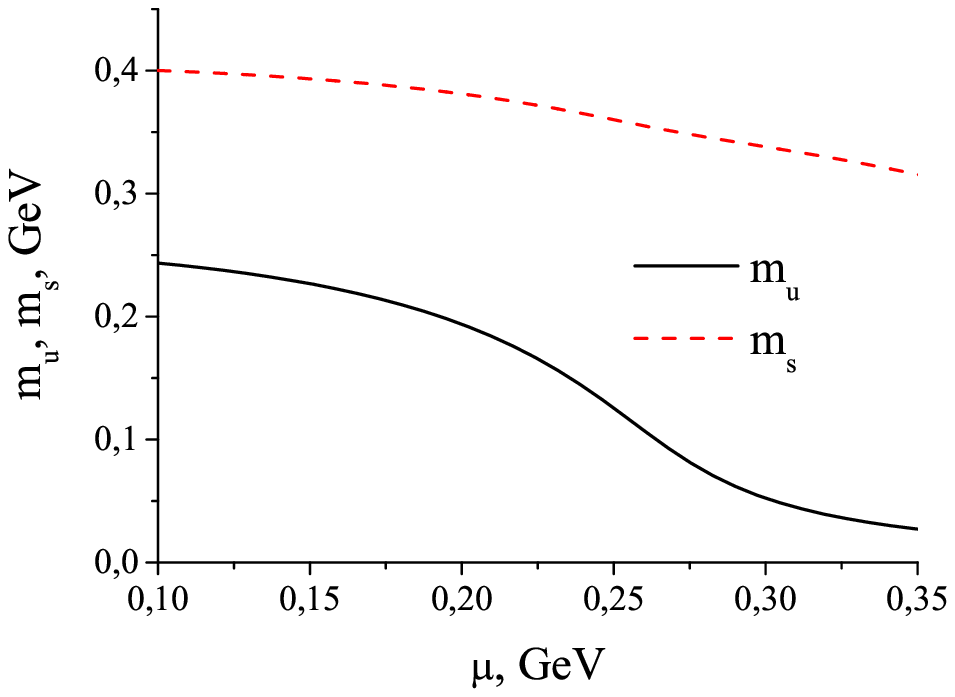}}
\end{center}
\end{figure}

Following the usual procedure of finite $T$ and $\mu$ calculations
in the NJL model, we make an assumption here that the model
parameters $G_1$, $G_2$, $K$, $m_{\rm u}^0$, $m_{\rm s}^0$ and
$\Lambda_3$ do not depend on $T$ and $\mu$ \cite{Ebert:1992ag}.
The dependence of the constituent quark masses $m_{\rm u}$,
$m_{\rm s}$ on $T$ and $\mu$ is found from gap equations
(\ref{gapu}) and (\ref{gaps}). One can see the behaviour of
$m_{\rm u}$ and  $m_{\rm s}$ {\it vs.} $\mu$ in figures~\ref{Fig_qmasses}
and \ref{Fig_qmasses_1} at $T=20$ and $100$~MeV.
As one can see from these plots, the mass of the strange quark changes little
with $\mu$ in the range of interest (in the hadron phase), therefore
$m_{\rm s}$ can be  considered as being approximately a constant fixed to its vacuum value.


\section{Vector-meson decays}

\begin{figure}[tbp]
\caption{The diagram corresponding to the $\rho(\omega,\phi) \to \gamma \gamma$ decay.
The crossed diagram is not
shown.}\label{Fig_rhogg}
\begin{center}
%{\includegraphics{rhotogg.eps}}
{\includegraphics{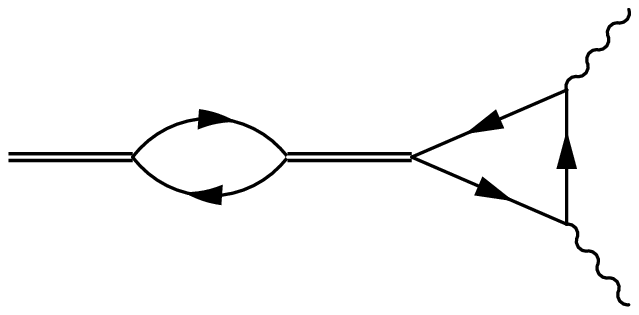}}
\end{center}
\end{figure}

\subsection{The decay $\rho\to\gamma\gamma$}
The direct decay of the neutral $\rho$-meson to two photons is
forbidden in vacuum, but in dense medium the transition of $\rho$
to the $a_0$-meson becomes possible because of the explicit
breaking of Lorentz symmetry for the ground state. The created
thereby state with quantum numbers of a scalar meson decays then
to a couple of photons. Schematically, the process is represented
by the diagram drawn in \fref{Fig_rhogg}. First, a quark loop
gives vector-scalar mixing, which is followed by an intermediate
scalar resonance and ends up by the triangle quark loop
describing a decay of the scalar meson to photons. The amplitude of
this process has the form
\begin{eqnarray}\label{eqA}
A^{\alpha \mu \nu}_{\rho \to \gamma \gamma} = J^\alpha_{\rho \to a_0} D_{a_0} T^{\mu\nu}_{a_0
\to \gamma \gamma},
\end{eqnarray}
where $J^\alpha_{\rho \to a_0}$ describes  $\rho$-$a_0$ mixing,
$D_{a_0}$ is the $a_0$-meson propagator, and $T^{\mu\nu}_{a_0 \to
\gamma \gamma}$ is the amplitude of the $a_0\to\gamma\gamma$ decay
(see \fref{Fig_rhogg}).
\begin{figure}[tbp]
\caption{The masses of $\pi$, $a_0$, $\sigma$, $\rho$, $\eta$ and
$\eta'$ -mesons as functions of $\mu$ at $T=20$~MeV  and
$T=100$~MeV.}\label{Fig_Mmasses}
\begin{center}
%{\includegraphics[scale=1]{MesT=20.eps}}
{\includegraphics[scale=0.97]{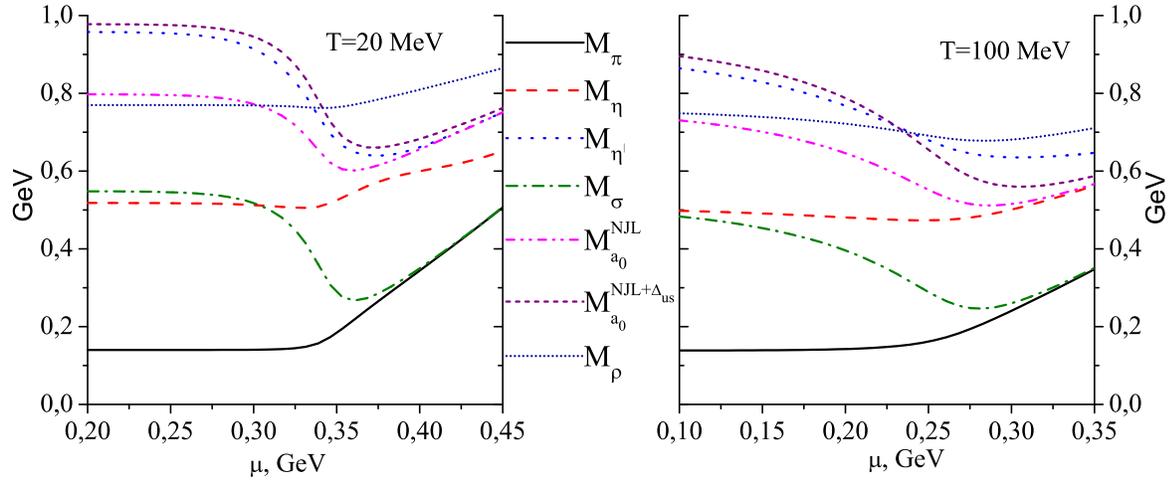}}
\end{center}
\end{figure}

Let us consider the right hand side of (\ref{eqA}) in detail.
The $\rho$-$a_0$ mixing is given by~(see~\cite{Blaschke:1997jj})
\begin{eqnarray}
J^\alpha_{\rho \to a_0}&= \frac{C_{\rho \to a_0}}{4}\int \frac{d^4 k}{(2\pi)^4}
\tr\left[S(k_-,T,\mu)\gamma^\alpha S(k_+,T,\mu)\right]\label{Cra0}\\
&= 2m_{\rm u}C_{\rho \to a_0}\int \frac{d^4 k}{(2\pi)^3}k^\alpha
\frac{\delta(k_+^2-m^2)}{k_-^2-m^2}
 \left(n(\mathbf{k}_+;T,\mu)-n(\mathbf{k}_+;T,-\mu)\right)\nonumber\\
&\times \left(\theta(k_+^0)-\theta(-k_+^0)\right),\nonumber\\
C_{\rho \to a_0}&=4 N_c g_{a0} g_\rho,\qquad
k_\pm=k\pm\frac{p}{2},\\
\quad g_{a_0}&=(4I_2^{\Lambda_3}(m_{\rm u},T,\mu))^{-1/2},\qquad
 g_\rho=\sqrt{6}g_{a_0},
\end{eqnarray}
where $g_{a_0}$ and $g_\rho$ are the $a_0$ and $\rho$ meson-quark coupling constants, $k$ is the internal
quark momentum, and $p$ is the four-momentum of the decaying meson.

In vacuum, at $\mu=T=0$, $J^\alpha_{\rho \to a_0}=0$, whereas
in medium it is nontrivial and vanishes only if the particle is at rest
in the heat-bath frame.
Moreover, $J^\alpha_{\rho \to a_0}$ satisfies gauge invariance:
$p_\alpha J^\alpha_{\rho \to a_0}=0$, which allows simplify the
following expressions, using the relation
\begin{eqnarray}
|J|^2=(J^\alpha_{\rho \to a_0})^*J_{\alpha,\rho \to
a_0}=-\frac{p^2}{{\mathbf{p}}^2}|J^0_{\rho \to a_0}|^2, \label{contr}
\end{eqnarray}
where $J^0_{\rho \to a_0}$ is computed by integration over angles in the heat-bath frame
\begin{eqnarray}
J^0_{\rho \to a_0}&=
\frac{C_{\rho \to a_0} m_{\rm u}}{(4\pi)^2|\mathbf{p}|}\!\!\int\limits_{m_{\rm u}}^\infty\!d E\,\delta(E)\!
 \left[ \left(2E+p^0\right) \ln\left(F_+\right) + \left(2E-p^0\right)  \ln\left(F_-\right)
		  \right],\label{J0} \\
 \delta(E)& =
 \frac{\mathrm{sinh}({\mu}/{T})}{\mathrm{cosh}({\mu}/{T})+\mathrm{cosh}({E}/{T})},\\
 F_\pm&=\frac{p^2\pm2p_0E+2|\mathbf{p}|q}{p^2\pm2p_0E-2|\mathbf{p}|q},
\end{eqnarray}
with $q=\sqrt{E^2-m^2}$.

The second multiplier in (\ref{eqA}) is the $a_0$-meson propagator calculated on
the mass-shell of the $\rho$-meson in medium,
\begin{eqnarray}
D_{a_0} = (M_{a_0}^2-M_\rho^2-i \Gamma_{a_0}(M_\rho)M_{a_0})^{-1}.
\end{eqnarray}
As is known from various investigations \cite{Ebert:1992ag} (see
also \fref{Fig_Mmasses}), the $\rho$-meson mass is a smooth
function  of $T$ and $\mu$ and slightly decreases when the
conditions of chiral symmetry restoration are being approached.
Therefore, for rough estimates it can be assumed to be constant in
the hadron phase.

As it follows from the NJL model calculations \cite{Ebert:1992ag},
the $\sigma$-meson mass drops down significantly with growing
temperature and chemical potential, until it becomes almost
degenerate with the pion mass when chiral symmetry is restored.
The same is expected for the $a_0$-meson. According to the NJL
model, the $a_0$ mass is to be found from the expression \cite{Volkov:1998ax}
\begin{eqnarray}
{M^2_{a_0}}=g^2_{a_0}\left[\frac{1}{G_{a_0}} - 8I^{\Lambda_3}_1(m_{\rm u})\right] + 4m^2_{\rm u}, \nonumber\\
G_{a_0}=G_1 - 4Km_{\rm s}I_1^{\Lambda_3}(m_{\rm s}),
\end{eqnarray}
which gives an underestimated value $M_{a_0}\approx 800$ MeV,
comparing to the measured mass $ M_{a_0}^{\mathrm{exp}} =
984.7\pm1.2$ MeV \cite{PDG}.

A possible explanation to the mass deficit in the NJL calculations
is the assumption of the existence of a four-quark component in
the $a_0$-meson
\cite{Jaffe:1976ig,Achasov:2003aa,Gerasimov:2004kq}, which is
usually ignored in the NJL-like models. To take into account the
four-quark component, we introduce an additional term $\Delta$ into
the mass formula for the $a_0$-meson
\begin{eqnarray}
{M_{a_0}^*}^2=M_{a_0}^2 + \Delta.
\end{eqnarray}
The model part of the mass of the $a_0$-meson as well as the order parameter $m_{\rm u}$
decrease with growing $T$ and $\mu$, as expected.
It is natural to assume that $\Delta$ has
a similar behaviour. We take $\Delta$ in the simplest form and consider two cases:
i) $\Delta=\Delta_{\rm u}$ and ii) $\Delta=\Delta_{\rm us}$ where
  $\Delta_{\rm u} = 4 m_{\rm u}^2$ and $\Delta_{\rm us}= 2.75 m_{\rm u} m_{\rm s}$
which is enough to reproduce the measured $a_0$ mass in vacuum.
After the phase transition, $\Delta$ vanishes, which is required
by restoration of chiral symmetry.

\begin{figure}[tbp]
\caption{Mixing angles for scalar and pseudoscalar mesons in the
NJL model for $T=20$ ~MeV and $T=120$~MeV as functions of chemical potential
$\mu$.}\label{Fig_angles}
\begin{center}
%{\includegraphics[scale=1]{anglesT=20.eps}}
{\includegraphics[scale=1]{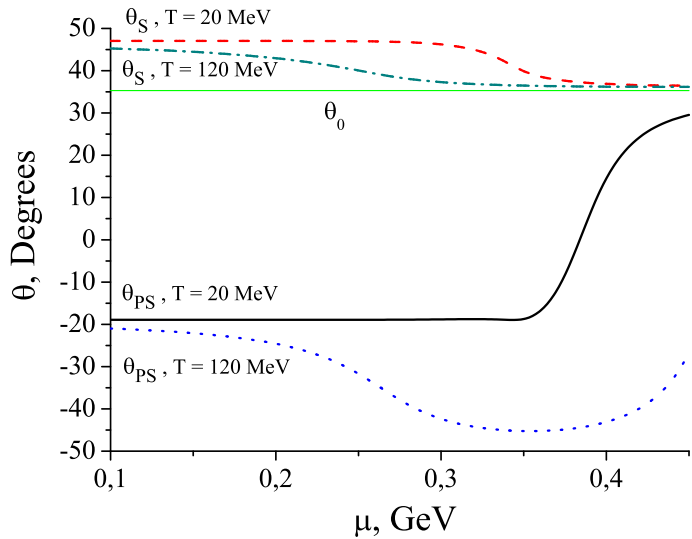}}
\end{center}
\end{figure}

\begin{figure}[p]
\caption{The widths of decays $a_0\to\eta\pi$ and
$\sigma\to\pi\pi$ in the NJL model for $T=20$~MeV  as functions of
chemical potential $\mu$.}\label{Fig_widths}
\begin{center}
\includegraphics[scale=1]{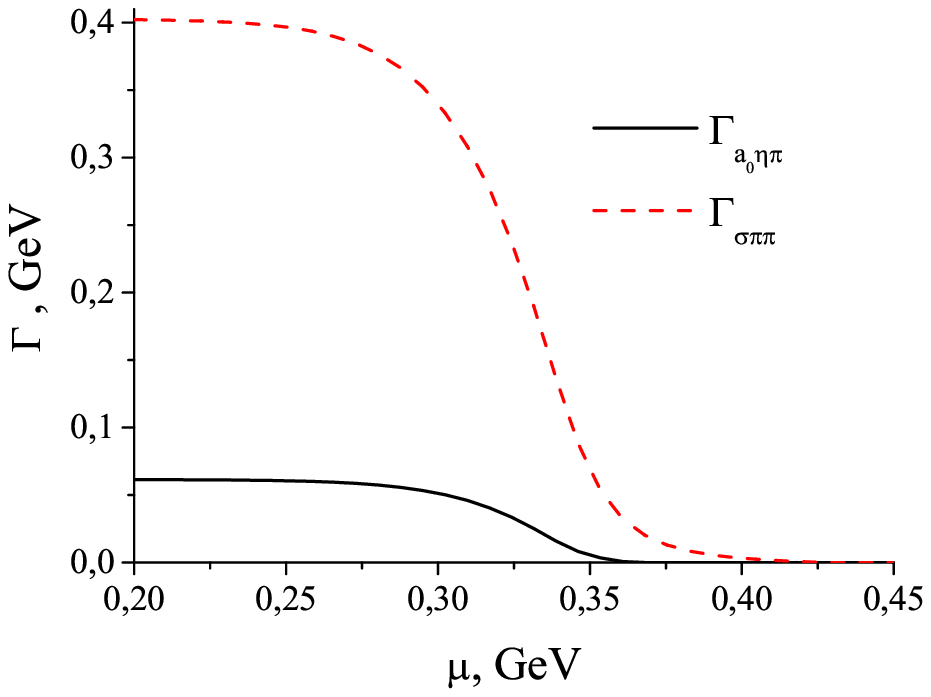}
\end{center}
\end{figure}

\begin{figure}[p]
\caption{The widths of decays $a_0\to\eta\pi$ and
$\sigma\to\pi\pi$ in the NJL model for  $T=120$~MeV  as functions of
chemical potential $\mu$.}\label{Fig_widths1}
\begin{center}
%{\includegraphics[scale=1]{wT=120.eps}}
{\includegraphics[scale=1]{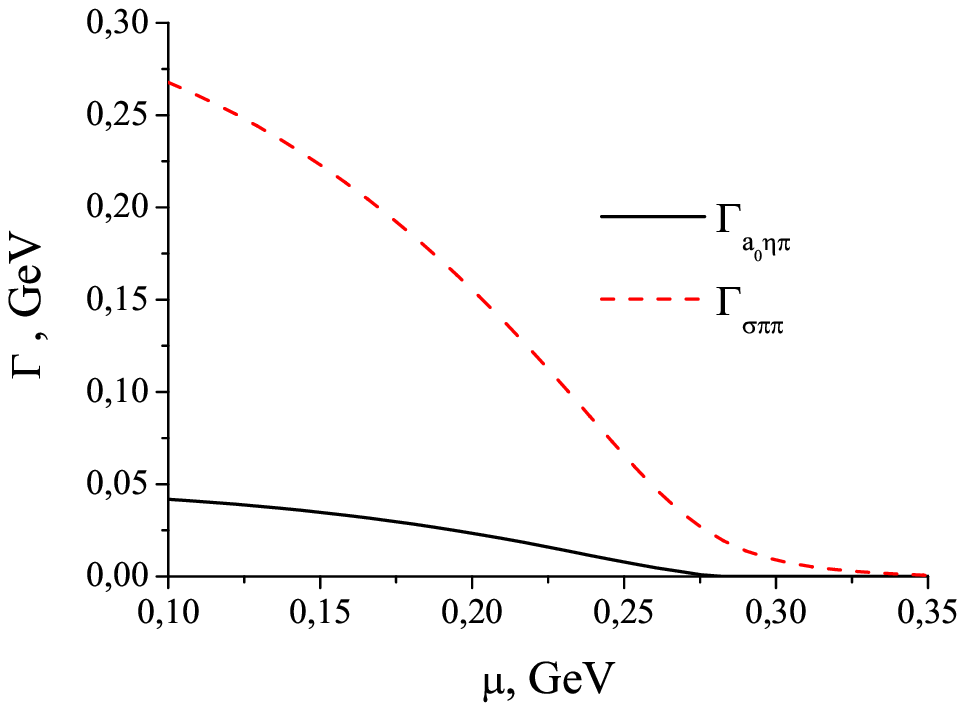}}
\end{center}
\end{figure}

The second quantity in the propagator of the $a_0$-meson is its width.
The partial decay mode $a_0 \to \eta \pi$ determines almost 100\% of the total width,
for which we obtain from the NJL model
\begin{eqnarray}
\Gamma_{a_0}(M_\rho)&\approx&\Gamma_{a_0\eta\pi}(M_\rho)=
\frac{g_{a_0\eta\pi}^2}{16 \pi M_{\rho}}
\sqrt{\left.1-\left[\frac{M_\eta+M_\pi}{M_\rho}\right]^2\right.}\nonumber\\
&\times&\sqrt{\left.1-\left[\frac{M_\eta-M_\pi}{M_\rho}\right]^2\right.},\\
\qquad g_{a_0\eta\pi}&=& 2m_{\rm u} g_{a_0} \sin\bar{\theta}_{\rm
P}, \qquad {\bar \theta}_{\rm P}=\theta_{\rm P} - \theta_0,
\end{eqnarray}
where $\theta_0 \approx 35.3^{\circ}$ is the ideal mixing angle
$({\rm ctg}~ \theta_0={\sqrt 2})$ and $\theta_{\rm P}$ is the
singlet-octet mixing angle for pseudoscalar mesons. The dependence
of the mixing angle on the  chemical potential $\mu$ is given in
\fref{Fig_angles} for $T=20$ and $T=120$ MeV.
%\begin{figure}[tbp]
%\caption{Mixing angles for scalar and pseudoscalar mesons in the
%NJL model for  T=120 MeV  as functions of chemical potential
%$\mu$.}\label{Fig_angles1}
%\begin{center}
%{\includegraphics[scale=1]{anglesT=121.eps}}
%{\includegraphics[scale=1]{figure6.eps}}
%\end{center}
%\end{figure}
The behaviour of $\Gamma_{a_0\eta\pi}(M_\rho)$ {\it vs.\/} $\mu$
is shown in figures~\ref{Fig_widths} and \ref{Fig_widths1}.

The last multiplier in (\ref{eqA}),
$T^{\mu\nu}_{a_0\to\gamma\gamma}$, is the amplitude describing the
two-photon decay of the $a_0$-meson. The expression for it is well
known \cite{Volkov:1997dx,Volkov:1994eu}
\begin{eqnarray}\label{a0tophotons}
T^{\mu\nu}_{a_0\to\gamma\gamma} =  C_{a_0\to \gamma
\gamma}f_1(T,\mu)F^{\mu\nu},\quad C_{a_0\to \gamma \gamma}=\frac{2
\alpha g_{a_0}}{3\pi m_{\rm u}},
\end{eqnarray}
where $\alpha=1/137$, $F^{\mu\nu}$ is the electromagnetic tensor, and function $f_1$ is defined as
\cite{Volkov:1997dx}
\begin{eqnarray} \label{f1}
f_1(T,\mu)&=&1-\frac{3}{2}m_{\rm u}^2(T, \mu)\int_0^\infty dk
\frac{k^3}{E^6(k)}
\ln\bigg[\frac{E(k)+k}{E(k)-k}\bigg]\nonumber\\
&\times& \big[ n(k;T,\mu)+n(k;T,-\mu)\big].
\end{eqnarray}

\begin{figure}[tbp]
\begin{center}
\caption{Two-photon decay width of $\rho$-meson with the $a_0$
mass corrected by $\Delta=\Delta_{\rm u}$  as a function of $\mu$
and $|\mathbf{p}|$ for $T=20$ MeV.}\label{Fig_Ma0}
%{\includegraphics[scale=1]{T20a0m.eps}}
{\includegraphics[scale=1]{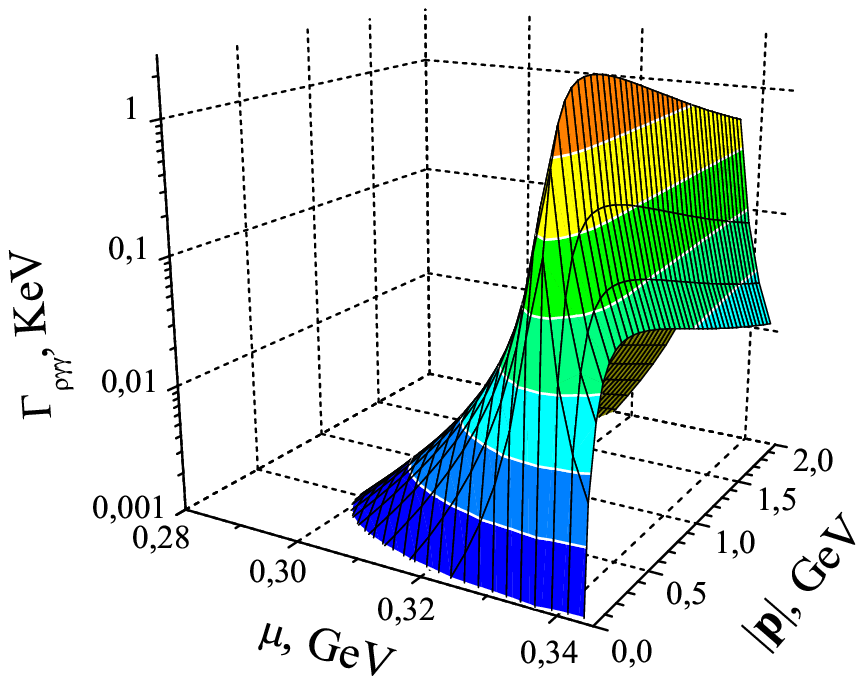}}
\end{center}
\end{figure}

\begin{figure}[tbp]
\begin{center}
\caption{Two-photon decay width of the $\rho$-meson with the $a_0$
mass corrected by $\Delta=\Delta_{\rm us}$  as a function of $\mu$
and $|\mathbf{p}|$ for $T=20$ MeV.}\label{Fig_Ma0_1}
%{\includegraphics[scale=1]{T20a0mums.eps}}
{\includegraphics[scale=1]{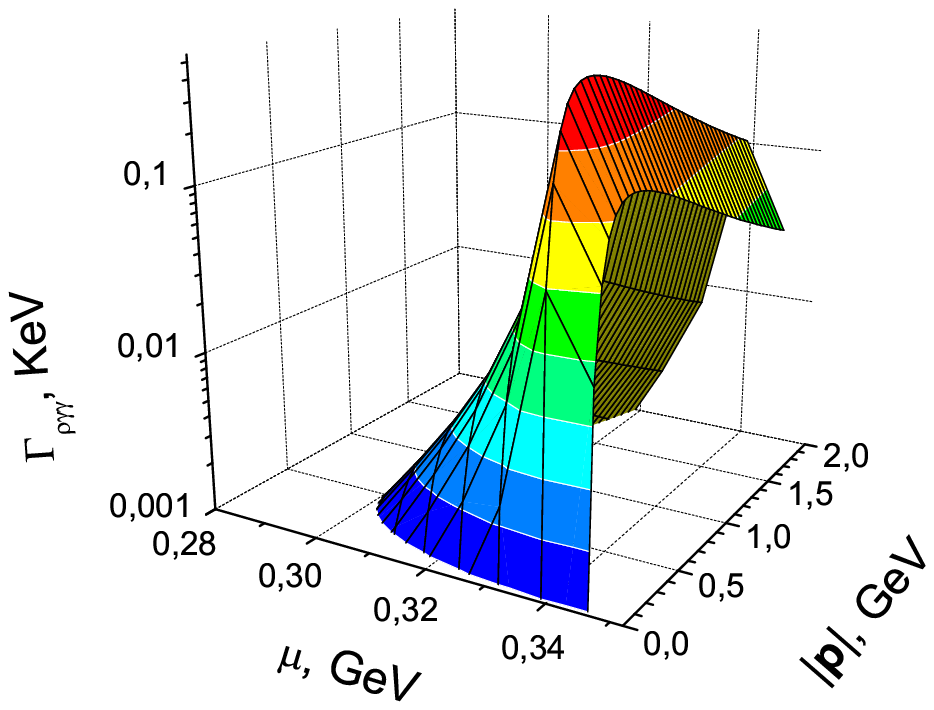}}
\end{center}
\end{figure}

\begin{figure}[tbp]
\caption{Two-photon decay width of the $\rho$-meson with the $a_0$
mass corrected by $\Delta=\Delta_{\rm u}$  as a function of $\mu$
and $|\mathbf{p}|$ for $T=120$ MeV.}\label{Fig_Ma0m}
\begin{center}
%{\includegraphics[scale=1]{T100a0m.eps}}
{\includegraphics[scale=1]{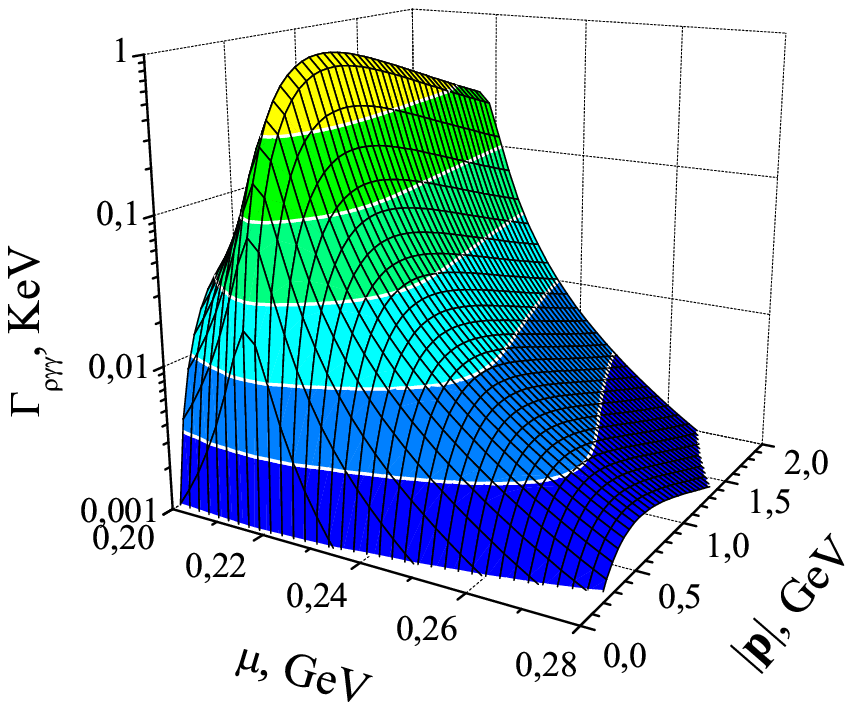}}
\end{center}
\end{figure}

\begin{figure}[tbp]
\caption{Two-photon decay width of the $\rho$-meson with the $a_0$
mass corrected by $\Delta=\Delta_{\rm us}$  as a function of $\mu$
and $|\mathbf{p}|$ for $T=120$ MeV.}\label{Fig_Ma0m_1}
\begin{center}
%{\includegraphics[scale=1]{T108a0mums.eps}}
{\includegraphics[scale=1]{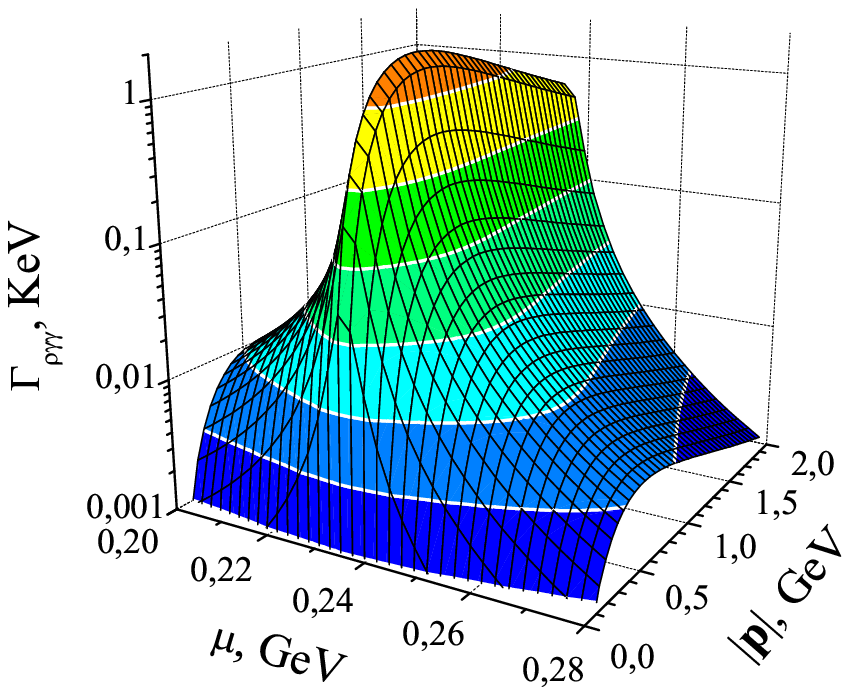}}
\end{center}
\end{figure}

After calculating the square of the absolute value of the amplitude,
summing over the polarization of the final state photons and
integrating over the solid angle, one obtains the width
\begin{equation}\label{Grho2gg}
\Gamma_{\rho \to \gamma \gamma} = \frac{\alpha^2 M_{\rho}^6 g_{a_0}^2 }{432
\pi^3 m_{\rm u}^2 E_\rho|\mathbf{p}|^2}\cdot\frac{|J_{\rho \to
a_0}|^2}{(M_{a_0}^2-M_\rho^2)^2+\Gamma_{a_0}^2(M_\rho)M_{a_0}^2},
\end{equation}
where $E_\rho=\sqrt{\mathbf{p}|^2+M_\rho^2}$.
Numerical estimates  for the width of the decay $\rho \to \gamma
\gamma$ as of a function of three-momentum $\mathbf{p}$ and
chemical potential $\mu$ at temperatures $T=20$ and $120$ MeV  are
shown in figures~\ref{Fig_Ma0}--\ref{Fig_Ma0m_1} with
$\Delta=\{\Delta_{\rm u},\Delta_{\rm us}\}$. It is easy to see
that at some values of $T$ and $\mu$ the two-photon decay width of
the $\rho$-meson can be comparable to the two-photon decay width
of scalar and pseudoscalar mesons in vacuum (see table
\ref{tabu}).
\begin{figure}[tbp]
\caption{Two-photon decay width of the $\omega$-meson as a function of
$\mu$ and $|\mathbf{p}|$ for $T=20$.}\label{Fig_Ms}
\begin{center}
%{\includegraphics[scale=1]{T20s.eps}}
{\includegraphics[scale=1]{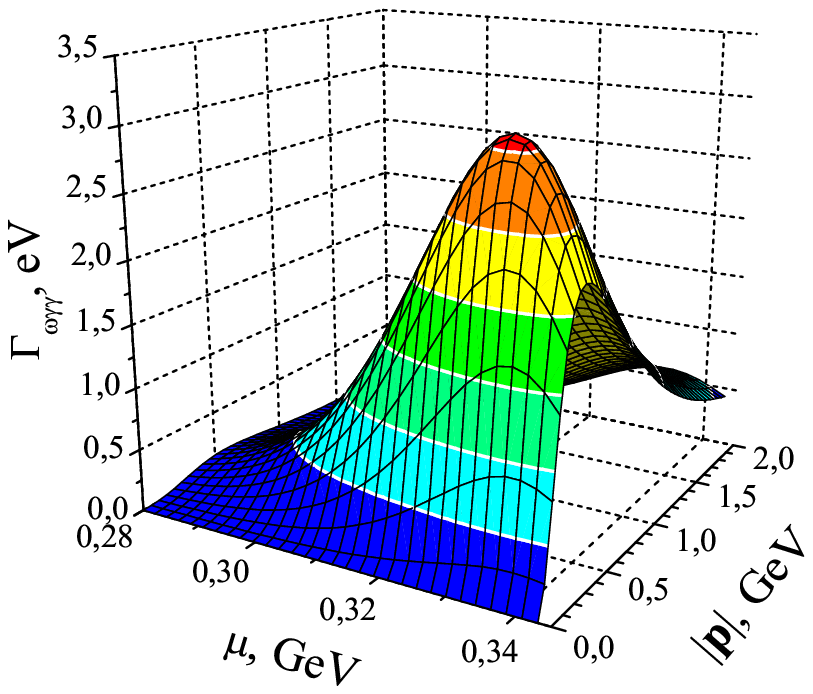}}
\end{center}
\end{figure}
\begin{figure}[tbp]
\caption{Two-photon decay width of the $\omega$-meson as a function of
$\mu$ and $|\mathbf{p}|$ for $T=120$~MeV.}\label{Fig_Ms_1}
\begin{center}
%{\includegraphics[scale=1]{T100s.eps}}
{\includegraphics[scale=1]{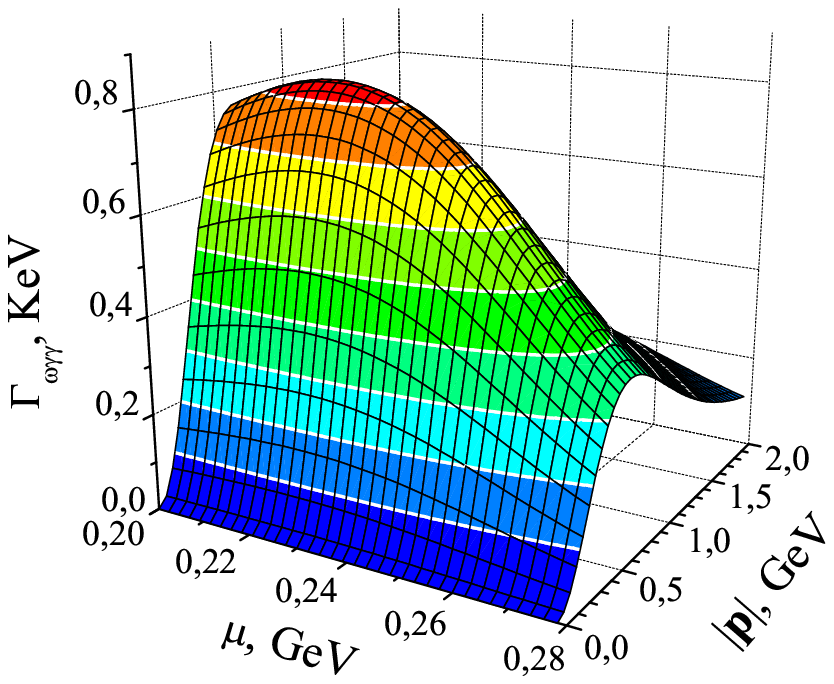}}
\end{center}
\end{figure}

Analogously,  the two-photon decay of the $\omega$-meson is given
by the same diagram as for $\rho$. The difference is that $\omega$
mixes with the isoscalar scalar meson $\sigma$. According to the
NJL model, the mass of the $\sigma$-meson is to be found from the
equation
\begin{eqnarray}\label{sigmaMass}
M^2_{\sigma}\approx
g_{a_0}^2\left(\frac{1}{G_\sigma}-8I_1^{\Lambda_3}(m_{\rm
u})\right)
+ 4m^2_u,\\
G_\sigma=G_1+4m_{\rm s} I_1^{\Lambda_3}(m_{\rm s}).
\end{eqnarray}
 The width of $\sigma$ is mostly
determined by its strong decay to two pions and is predicted by
the NJL model (on the mass-shell of the $\omega$-meson) to be of
the form
\begin{eqnarray}
\Gamma_{\sigma}(M_\omega)&\approx&\Gamma_{\sigma \pi
\pi}(M_\omega)= \frac{3g_{\sigma \pi \pi}^2}{32 \pi
M_\omega}\sqrt{1-\frac{4M_\pi^2}{M_\omega^2}},\\
\qquad g_{\sigma \pi \pi}&=& 2m_{\rm u} g_{a_0}
\cos\bar{\theta}_{\rm S}.
\end{eqnarray}
The amplitude for the decay $\omega\to\gamma\gamma$ can be
obtained from (\ref{eqA}) simply by replacing the factor
$C_{a_0\to \gamma \gamma}$ in $T_{a_0\to\gamma\gamma}^{\mu\nu}$
(see (\ref{a0tophotons})) by $C_{\sigma\to \gamma \gamma}=10
\alpha g_{a_0}/(9\pi m_{\rm u})$ and by using the $\sigma$-meson
propagator instead of $a_0$ $(D_{a_0}\to D_{\sigma})$
\begin{eqnarray}
D_{\sigma} = (M_{\sigma}^2-M_\omega^2-i
\Gamma_{\sigma}(M_\omega)M_{\sigma})^{-1}.
\end{eqnarray}
The angle $\bar{\theta}_{\rm S}$ is defined as the difference
between the mixing angle for scalar mesons $\theta_{\rm S}$ and
the ideal mixing~$\theta_0$: $\bar{\theta}_{\rm S}=\theta_{\rm
S}-\theta_0$. Its  behaviour with respect to $\mu$ is given in
\fref{Fig_angles}, from which one can
see that the singlet-octet mixing among scalar mesons is almost
ideal and we put $\bar\theta_S= 0$ in further calculations.

A peculiarity of the decay $\rho \to \gamma \gamma$ is that the
$a_0$-meson mass is larger than the $\rho$-meson mass in vacuum
and significantly drops down when approaching the transition to
the phase with restored chiral symmetry. One can conclude from
this that in particular conditions the masses of $a_0$ and $\rho$
become degenerate, which can be followed by a resonant
amplification of the decay $\rho\to\gamma\gamma$ and noticeable
enhancement near the $\rho$-meson mass in the two-photon spectrum.
As to the $\sigma$-meson, it is always lighter  than $\omega$.
Moreover, their mass difference grows with $\mu$, and, as a consequence,  no
resonant phenomenon occurs in this decay. Therefore,  the number
of two-photon events related to decays of vector meson will be
dominated by those coming from
 $\rho \to \gamma \gamma$ rather than from $\omega \to \gamma \gamma$,
as one can see from figures \ref{Fig_Ms} and \ref{Fig_Ms_1}.

\subsection{The decay $\phi\to\gamma\gamma$}\label{subsec:phi2gg}
The decay of the $\phi$-meson occurs due to $\phi$-$f_0$ mixing
($\phi\to f_0(980)\to\gamma\gamma$), where the photon pair is
produced by the decay of the intermediate $f_0(980)$-meson. As one can
see from \fref{Fig_angles}, the
mixing among scalar mesons is almost ideal, and one can consider
the
 $f_0(980)$-meson as composed of $s$- and $\bar s$-quarks only. The $\phi$-meson
also consists  mostly of strange quarks. Using this
approximation together with the fact that the mass of the strange
quark is almost constant (see \fref{Fig_qmasses} and
\ref{Fig_qmasses_1}) in the hadron phase, one can roughly estimate
the decay of $\phi$ to $\gamma\gamma$.

 The $\phi$-$f_0(980)$ mixing is represented by
an integral of the form \eref{J0}, where one one should replace
the $u$-quark mass by the mass of the strange quark ($m_{\rm u}\to
m_{\rm s}$, $J^0_{\rho\to \omega}\to J^0_{\phi\to f_0}$).
Qualitatively, the decay of the $\phi$ meson to photons has the
behaviour with respect to $\mu$ similar to that  of the lighter
vector mesons ($\rho$ and $\omega$), so we omit here the plots
for the $\phi\to\gamma\gamma$ decay rate. We should like only to note here that the width
for the decay $\phi\to\gamma\gamma$ can be as large as 2~keV at
appropriate conditions in the hadron phase at the three-momentum
about 1~GeV. The width is greater for larger momenta: at
$|\mathbf{p}|\sim 5$~GeV the width can reach 8~keV, but such
momenta are less probable at temperatures of order 100--200 MeV.

\subsection{Strong decays $\rho \to \pi \eta$ and $\omega \to \pi \pi$}

\begin{figure}[tbp]
\caption{Strong decay width $\rho \to \pi \eta$ at  $T=20$
MeV.}\label{Fig_strong}
\begin{center}
%{\includegraphics[scale=1]{rhopieta20.eps}}
{\includegraphics[scale=1]{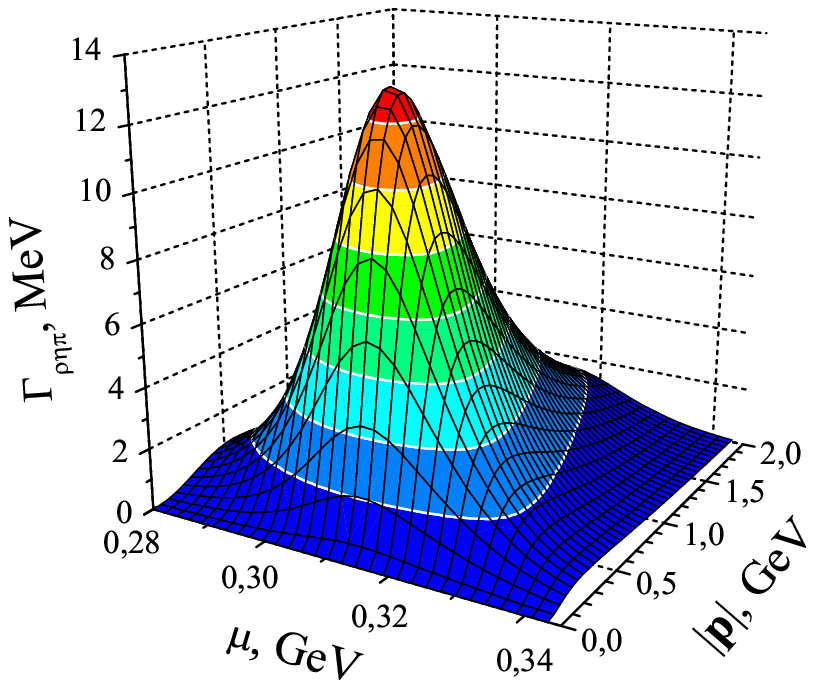}}
\end{center}
\end{figure}

\begin{figure}[tbp]
\caption{Strong decay width  $\omega \to \pi \pi$ at $T=20$
MeV.}\label{Fig_strong_1}
\begin{center}
%{\includegraphics[scale=1]{omegapipi20.eps}}
{\includegraphics[scale=1]{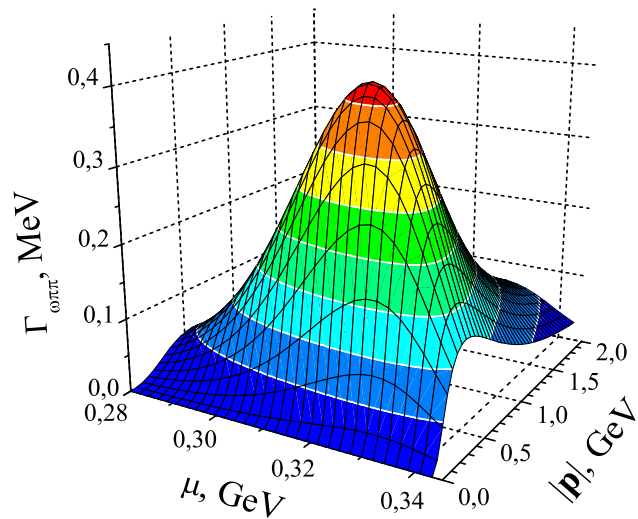}}
\end{center}
\end{figure}

The strong decays $\rho \to \pi \eta$ and $\omega \to \pi
\pi$\footnote{The in-medium decay width $\omega\to\pi\pi$  is
calculated for nuclear matter within a hadronic model including
mesons, nucleons and $\Delta$-isobars in
\cite{Broniowski:1999ji}.} can be considered in the same manner as
the electromagnetic ones which were discussed above. The difference  is that
an intermediate scalar meson decays to hadrons instead of photons
and one needs only to replace the third multiplier in (\ref{eqA})
by an appropriate amplitude for the corresponding strong decay.
 For these decay widths at temperature
$T=20$ MeV, we give numerical estimates
 in figures~\ref{Fig_strong} and \ref{Fig_strong_1}.
The maximal value of the width of the decay $\rho\to\gamma\gamma$
is about 40~MeV, which is comparable with the main decay of the
$\rho$-meson --- the decay to two pions (150~MeV). The decay
$\omega\to\pi\pi$ can occur in vacuum due to the broken isotopic
invariance ($m_{\rm u}\not=m_{\rm d}$) and gives
$\Gamma_{\omega\to\pi\pi}=145\pm25$ keV; in medium, its decay rate
grows significantly and reaches $1.5$~MeV.
  One should note here that in dense medium
these strong decays become comparable with regular electromagnetic
decays of scalar and pseudoscalar mesons whose widths are shown in
\tref{tabu}.


\begin{table}
\caption{Two-photon decays of scalar and pseudoscalar mesons in vacuum
\cite{PDG}.}\label{tabu}
\begin{indented}
\item[]\begin{tabular}{lllllll}
\br
Particle  & Mass, MeV&$\Gamma_{\gamma \gamma}$, keV  \\
\mr
$\pi^0$ &$134.9766\pm0.0006$& $(7.8\pm0.5)\cdot 10^{-3}$ \\

$\eta$&$547.75\pm0.12$&$1.29\pm0.07$ \\

$\eta^\prime$&$957.78\pm0.14$&$4.29\pm0.15$ \\

$\sigma$ &$400-1200$& $\sim 1$ \\

$a_0$&$984.7\pm1.2$&$0.3\pm0.1$ \\

$f_0$&$980\pm10$&$0.39^{+0.1}_{-0.13}$ \\
\br
\end{tabular}
\end{indented}
\end{table}

\section{Scalar meson decays}
\subsection{The decay $\sigma\to e^+ e^-$}

\begin{figure}[tbp]
\caption{The diagram that describes dilepton decays of scalar meson in medium via
scalar-vector mixing}\label{Fig_s2ee}
\begin{center}
\includegraphics{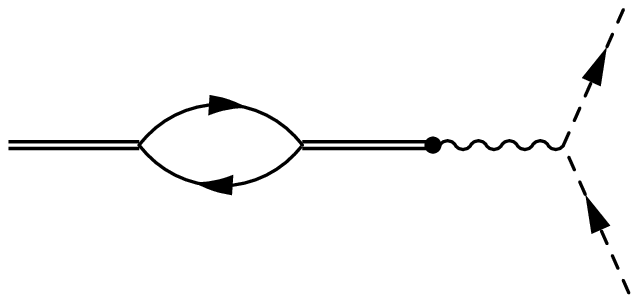}
\end{center}
\end{figure}

The direct decay of the $\sigma$-meson to dileptons  is also
forbidden in vacuum and is allowed in dense medium due to
$\sigma$-$\omega$ mixing. After transition of $\sigma$ to $\omega$,
the latter transforms to a photon (according to VMD) which then
produces an electron and a positron (see the diagram in \fref{Fig_s2ee}).
The corresponding amplitude is
\begin{eqnarray}\label{eqB}
A^{s_1 s_2}_{\sigma\to e^+e^=} = J^\alpha_{\sigma \to \omega}
D_{\omega\, \alpha\beta} T^{\beta\, s_1 s_2}_{\omega \to e^+e^-},
\end{eqnarray}
where $J^\alpha_{\sigma \to \omega}$ describes the
$\sigma$-$\omega$ mixing , $D_{\omega\,\alpha\beta}$ is the
$\omega$-meson propagator, and $T^{\beta\, s_1 s_2}_{\omega\to
e^+\e^-}$ is the amplitude of the decay $\omega\to e^+e^-$.
Fortunately, the $\sigma$-$\omega$ mixing coincides with
$\rho$-$a_0$ mixing
\begin{equation}
J^\alpha_{\sigma \to \omega}=J^\alpha_{\rho \to a_0},
\end{equation}
and the results of previous
section can be used here.
One then needs the $\omega$-meson propagator on the mass of the $\sigma$-meson in medium
$(p^2=M_\sigma^2)$:
\begin{eqnarray}\label{Domega}
D_{\omega\,\alpha\beta} = \frac{g_{\alpha\beta}-p_{\alpha}p_{\beta}/M_{\omega}^2}{
M_{\omega}^2-M_\sigma^2-i
\Gamma_{\omega}(M_\sigma)M_{\omega}}
\equiv\left(g_{\alpha\beta}-\frac{p_{\alpha}p_{\beta}}{M_{\omega}^2}\right) D_\omega.
\end{eqnarray}
The width of the $\omega$-meson is negligibly small and can be
neglected here because the mass difference for the $\omega$ and
$\sigma$ mesons increases while approaching the phase transition
and dominates in the denominator in (\ref{Domega}):
\begin{eqnarray}\label{Domega1}
D_{\omega} \approx \frac{1}{M_{\omega}^2-M_\sigma^2}.
\end{eqnarray}
For the same reason that was discussed in previous section, the
$\omega$-meson mass can be assumed to be constant. The
$\sigma$-meson mass (in the NJL model) has been already given in
(\ref{sigmaMass}) and is estimated to be about $M_{\sigma}\approx
550$ MeV in vacuum (see \fref{Fig_Mmasses}).

The last multiplier in the amplitude, $T^{\beta\, s_1 s_2}_{\omega\to
s_1 s_2 }$, is the amplitude describing the decay $\omega\to\gamma^*\to
e^+e^-$ in the VMD model:
\begin{eqnarray}\label{sig2ee}
T^{\beta s_1 s_2}_{\omega\to e^+e^-} = \frac{4\pi\alpha
M_\omega^2}{3g_\omega M_\sigma^2}  \bar
v(\mathbf{k}_2|s_2)\gamma_\beta u(\mathbf{k}_1|s_1),
\end{eqnarray}
where  $u(\mathbf{k}_1|s_1)$
and $\bar v(\mathbf{k}_2|s_2)$ are spinors corresponding to
the produced electron and positron, with $s_i (i=1,2)$ being their
 spin projections and $k_i$  their four-momenta.

\begin{figure}[tbp]
\caption{The decay $\sigma\to e^+e^-$ at $T=20$~MeV}
\label{Fig_sig2ee}
\begin{center}
\includegraphics{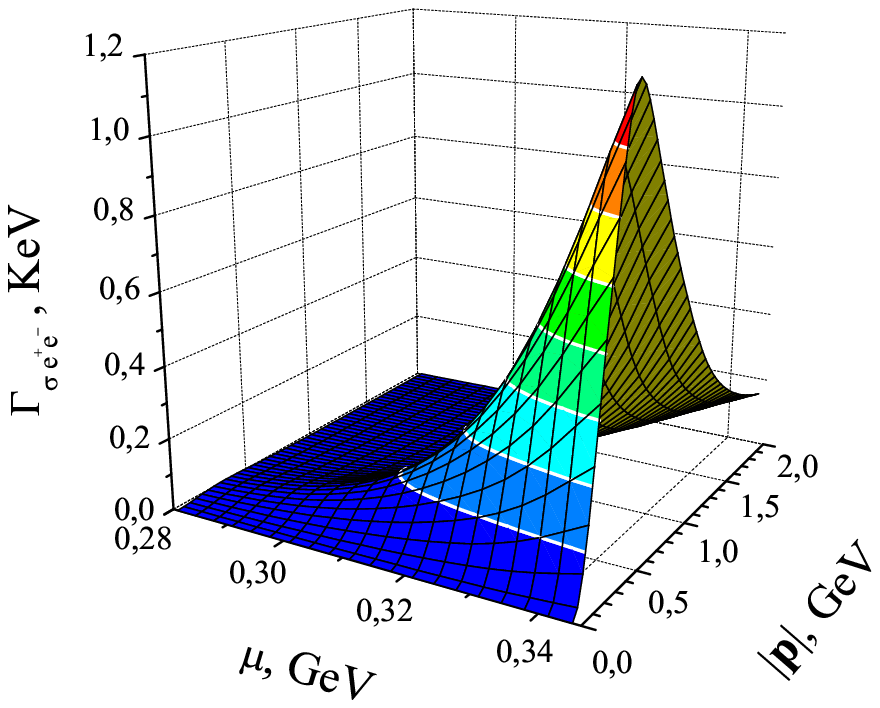}
\end{center}
\end{figure}

\begin{figure}[tbp]
\caption{The decay $\sigma\to e^+e^-$ at $T=120$~MeV}
\label{Fig_sig2ee_1}
\begin{center}
\includegraphics{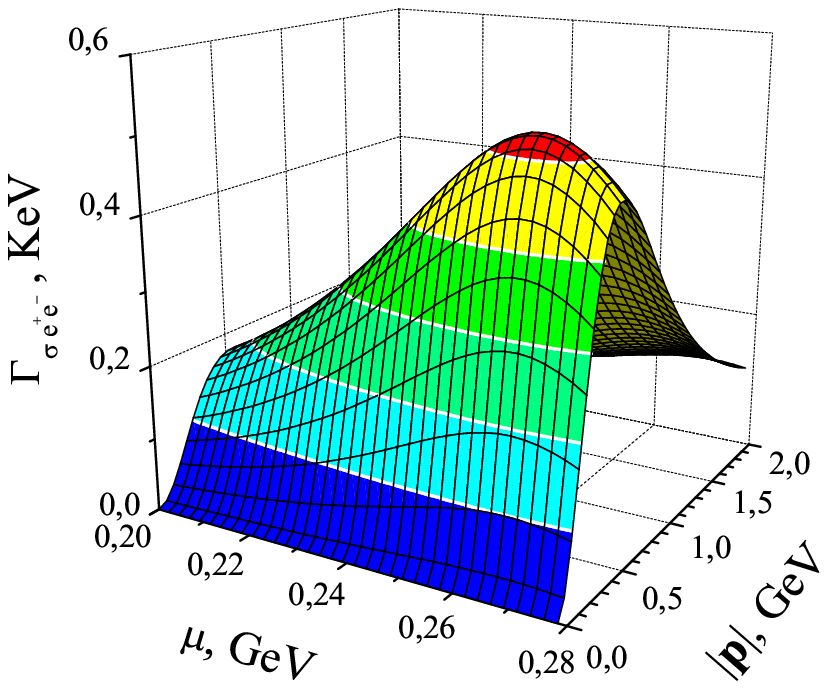}
\end{center}
\end{figure}

To calculate the partial width for the dilepton decay of  a scalar meson,
one needs then to take the square of the absolute value of the amplitude
(\ref{eqB}), sum over the spin projections (because the polarization of
leptons is not measured) and integrate over the solid angle in the
final state.
The result is
\begin{equation}
\Gamma_{\sigma\to e^+e^-}(M_\sigma,\mathbf{p})\approx
  \frac{4\alpha^2\pi^3 M_\omega^4}{27g_\omega^2 M_\sigma |\mathbf{p}|^2}
\frac{|J^0_{\rho\to a_0}|^2}{(M_\sigma^2-M_\omega^2)^2}.
\end{equation}
Numerically (see figures~\ref{Fig_sig2ee} and \ref{Fig_sig2ee_1}),
the decay $\sigma\to e^+e^-$ reaches 1.5~keV in
maximum.

\subsection{The decay $a_0\to e^+ e^-$}
\begin{figure}[tbp]
\caption{The decay $a_0\to e^+e^-$ at $T=20$~MeV}
\label{Fig_a02ee}
\begin{center}
\includegraphics{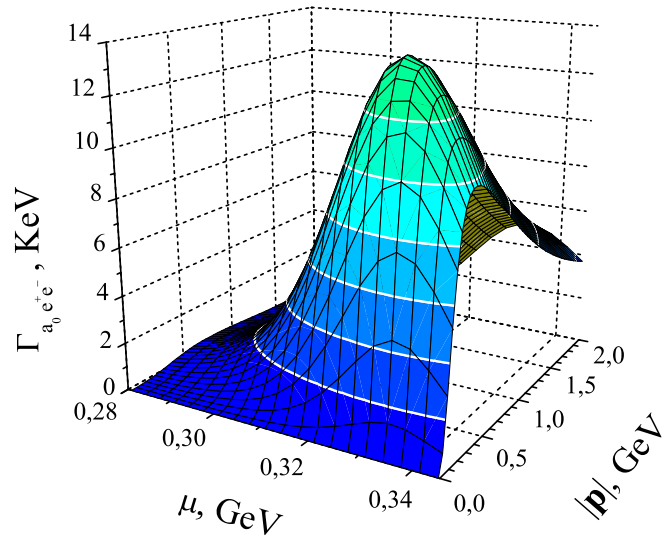}
\end{center}
\end{figure}

\begin{figure}[tbp]
\caption{The decay $a_0\to e^+e^-$ at $T=120$~MeV}
\label{Fig_a02ee_1}
\begin{center}
\includegraphics{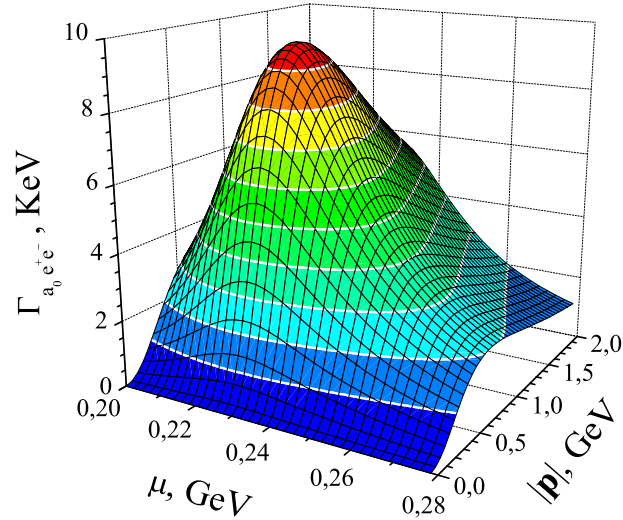}
\end{center}
\end{figure}

The decay  $a_0\to e^+e^-$ is qualitatively the same as $\sigma\to
e^+e^- $. The width is as follows
\begin{equation}
\Gamma_{a_0\to e^+e^-}(M_{a_0},\mathbf{p})\approx
  \frac{4\alpha^2\pi^3 M_\rho^4 }{3 g_\omega^2 M_{a_0} |\mathbf{p}|^2}
\frac{|J^0_{\rho\to a_0}|^2}{(M_{a_0}^2-M_\rho^2)^2+M_\rho^2
\Gamma_\rho(M_{a_0})^2}.
\end{equation}
The width of the $\rho$-meson is given by its decay to pions:
\begin{equation}
\Gamma_\rho(M_{a_0})\approx
\Gamma_{\rho\to\pi\pi}(M_{a_0})=\frac{g_\rho^2}{48\pi M_{a_0}^2} (
M_{a_0}^2-4M_\pi^2)^{3/2}.
\end{equation}
 The
numerical estimates for the decay $a_0\to\gamma\gamma$ are given
in figures \ref{Fig_a02ee} and \ref{Fig_a02ee_1}.
Due to the mass degeneration of $\rho$ and $a_0$
near the phase transition, a resonant enhancement is to be
observed in the decay $a_0\to e^+ e^-$. However, because of the
large width of the $\rho$-meson, there is no sharp resonance.
Nevertheless, insofar as the $\rho$-$\gamma$ transition
is three times greater than the $\omega$-$\gamma$ transition
(according to VMD) the rate of the decay $a_0\to e^+ e^-$ turns to
be larger almost by an order ($\sim 10$~keV at the maximum),
comparing to the decay $\sigma\to e^+ e^-$.

\subsection{The decay $f_0(980)\to e^+ e^-$}
As it has been already mentioned in section~3
 the mixing among scalar mesons is
almost ideal near the phase transition, and the
 $f_0(980)$-meson is considered as composed of $s$- and $\bar s$-quarks only.
For the $f_0(980)$ and $\phi$ masses assumed to be constant in the
hadron phase, one can roughly estimate the decay of $f_0(980)$ to
$e^+e^-$. This decay is mediated by the $\phi$-meson. The
$\phi$-$f_0(980)$ mixing is again represented by the integral of the
form \eref{J0} with the $u$-quark mass replaced by the mass of the
strange quark (see section 3). The other difference is
that the transition of the $\phi$-meson to a photon is larger that
that for the $\omega$-meson by factor $\sqrt{2}$ \cite{Volkov:1986zb}. Taking into
account all the written above, one obtains for $f_0\to e^+e^-$:
\begin{equation}
\Gamma_{f_0\to e^+e^-}(M_{f_0},\mathbf{p})\approx
  \frac{8\alpha^2\pi^3 M_\phi^4 }{27 g_\phi^2 M_{f_0} |\mathbf{p}|^2}
\cdot\frac{|J^0_{\phi\to f_0}|^2}{(M_{f_0}^2-M_\phi^2)^2+M_\phi^2
\Gamma_\phi(M_{f_0})^2}.
\end{equation}
Qualitatively, the behaviour of this width with respect to
temperature and chemical potential is similar to the decay
$\sigma(a_0)\to e^+ e^-$, and we do not show plots for this decay.
We should like to note here only that below $\mu=350$~MeV, the
maximal width is about 10~keV.

\section{Conclusion}

Two-photon decays of vector mesons and dilepton decays of scalar
mesons  are forbidden in free space but in medium they give an
additional contribution to the corresponding two-photon and
dilepton spectra in heavy-ion collisions. These decays are
determined by $\sigma$-$\omega$,  $\rho$-$a_0$ and
$\phi$-$f_0(980)$ mixing which strongly depend on temperature,
chemical potential and on the momentum of the decaying particle
in the medium. The mixing is mostly density-driven, and the effect
is to be observable for relatively high chemical potentials and
not too large temperatures. The maximum of the mixing is reached
for the those particles whose momentum corresponds to the
maximum in the momentum distribution at fixed $T$ and $\mu$. The
rate of each decay vanishes in free space and is maximal near the
transition from the hadron phase to the phase with restored chiral
symmetry. The conditions in which the effect is noticeable
correspond to a wide range in the phase diagram. The latter is
important for heavy-ion collisions because one can thereby hope
that the evolution of a fireball will lie mostly in this range.
Additional enhancement is produced by the resonant effect if the
$\rho$ in involved. This enhancement is similar to that in the
process $\pi\pi\to\gamma\gamma$ \cite{Volkov:1997dx,Volkov:2002bv}
and $\pi\pi\to\pi\pi$ \cite{Hatsuda:2001da}.

 The case of $\rho$-$a_0$ mixing seems
more intriguing than the others because there are points in the
phase diagram that correspond to conditions at which the masses of
$\rho$ and $a_0$ are close to each other. This leads to additional
amplification of the two-photon decay of the $\rho$-meson because
the intermediate $a_0$ meson becomes a sharp resonance. As a
result, the forbidden in vacuum decay $\rho\to\gamma\gamma$ turns
out to be comparable  (about 2.5 keV) to regular decays
$a_0\to\gamma\gamma$ (0.3 keV, see \tref{tabu}),
$\sigma\to\gamma\gamma$ ($1\div10$~keV) and $\eta\to\gamma\gamma$
($\sim 1.3$ keV). Similar amplification will also be observed in
the (strong) decay $\rho\to\eta\pi$, which reaches the maximum
value about 40 MeV. As to the $\omega$-meson, its decay to pions
significantly increases near the phase transition, comparing to its
value in vacuum.

Despite the degeneration of $\rho$ and $a_0$ masses, there is no
enhancement in the decay $a_0\to\gamma\gamma$ similar to that in
the decay $\rho\to\gamma\gamma$ because of the large width of the
$\rho$-meson. Concerning the decay $\sigma\to e^+e^-$, insofar as
$\sigma$-$\omega$ mass difference is relatively large and grows
near the phase transition, no resonance is observed here.

All the effects discussed here take place in the case of the
decays $\phi\to\gamma\gamma$ and $f_0(980)\to e^+e^-$.

In addition, we would like to remark that, according to VMD,
$\rho$-$\gamma$ mixing is three times larger than
$\omega$-$\gamma$. As a consequence, the number of dileptons
produced by $a_0$ exceeds that produced by $\sigma$ by an order.

In this paper we did not consider diagrams excluding intermediate
resonances, such as the decay of $\rho$-meson into a photon pair
through the triangle quark diagram (see
\cite{Teryaev:1996dv,Skalozub:1999iw}). Concerning the pion-loop
contribution, we do not consider them here because they give
next-to-leading order in the $1/N_c$ expansion. Concluding our
paper, we would like to emphasize that the processes discussed
here occur only in dense matter and can serve as indicators of
approaching the quark-gluon plasma in heavy-ion collision
experiments.

\ack The authors should like to thank S.~B.~Gerasimov,
Yu.~L.~Kalinovsky, A.~S.~Sorin and O.~V.~Teryaev for useful
discussions. The work is partially supported by RFBR Grant no.
05-02-16699 and ``Dynasty'' Foundation.

\end{document}